\documentclass[twocolumn,amsmath,showpacs,graphicx,amssymb]{revtex4}
\usepackage{amssymb}
\usepackage{amsmath}
\usepackage{graphicx}

\begin{document}

\title{Giant flexoelectric effect in two-dimensional boron-nitride sheets}
\author{Ivan Naumov$^1$, Alexander M. Bratkovsky$^1$ and V. Ranjan$^2$}
\affiliation{$^1$Hewlett-Packard Laboratories, Palo Alto, CA 94304, USA, $^2$North
Carolina State University, Raleigh, NC 27606, USA}
\date{\today }

\begin{abstract}
We find, with the use of first-principles calculations, that a
single-atom-thick boron-nitride (BN) sheet exhibits an unusual \textit{%
nonlinear} electromechanical effect:\ it becomes macroscopically polarized
when bent out-of-plane. The direction of the induced polarization is in the
plane of the film and it depends non-analitically on the corrugation wave
vector $\boldsymbol{k}$. The magnitude of the polarization can reach very
high values in spite of being at least quadratic in atomic displacements due
to BN sheets being able to tolerate large mechanical strains. The discovered
effect can find many applications, in particular, \ in a new type of
efficient and reliable nanogenerators.
\end{abstract}

\maketitle

Three-dimensional (3D) bulk crystals can generate a voltage either in
response to a mechanical strain $\partial _{i}u_{l}$ (piezoelectric effect
\cite{martin, tagantsev}) or to a strain gradient $\partial _{i}\partial
_{j}u_{l}$ (flexoelectric effect \cite{tagantsev,cross}), where $\boldsymbol{%
u}\left( \boldsymbol{r}\right) $ is the displacement vector, with $l$ its
Cartesian index, $\partial _{i}$ the gradient operator. The flexoelectric
effect is usually small and evades experimental detection unless large
strain gradients are externally imposed or artificially designed
inhomogeneous metamaterials are used \cite{cross, sharma}). It is commonly
assumed that the effects \textit{nonlinear} in $\partial _{i}u_{l}$ and $%
\partial _{i}\partial _{j}u_{l}$ on a polarization are negligible in bulk
dielectrics.

Contrary to 3D systems, $sp^{2}$-bonded 2D crystals, like graphene and boron
nitride (BN) \cite{saito, meyer}, are able to sustain much larger structural
distortions and, thus, exhibit new forms of electromechanical coupling. The
BN sheet, for example, becomes pyroelectric when it is wrapped into a chiral
or zigzag nanotube with the macroscopic polarization inversely proportional
to the inverse square of the tube radius, $1/R^{2}$, and directed along the
tube \cite{mele,nakhamson,sai}. Formally, this effect can be considered as a
\textit{quadratic} flexoelectric effect since $1/R^{2}$ is only a particular
form of $\left( \partial _{i}\partial _{j}u_{k}\right) ^{2}$.

Here, we predict another unusual \textit{nonlinear} electromechanical effect
in 2D BN monolayer that has not been noticed so far: generation of a
macroscopic \textit{in-plane} polarization in response to out-of-plane
atomic displacements like\ the corrugation $u_{z}(\boldsymbol{r})=A\sin (%
\boldsymbol{k\cdot r}+\varphi )$, where $\boldsymbol{k}$ is the undulation
wave vector, $\boldsymbol{r}$ is the in-plane vector, $z-$axis is
perpendicular to the plane, and $\varphi $ is some phase. Although such
displacements have \textit{no} in-plane components and produce \textit{zero}
net curvature, they nevertheless induce an in-plane macroscopic polarization
$\boldsymbol{P}\left( \boldsymbol{k}\right) \boldsymbol{,}$ which is purely
electronic in origin and related to the shifts of $\pi $ and $\sigma $
chemical bonds. It is remarkable that the induced polarization practically
does not depend on the phase $\varphi $ but strongly depends on wave vector $%
\boldsymbol{k}$, and there are some (\textquotedblleft optimal") directions
along which the absolute value $|\boldsymbol{P(k)}|$ goes through a maximum.
We show further that being at least quadratic in amplitude of atomic
displacements $A$, the corrugation-induced polarization is decomposed
basically into a sum of two contributions associated with the quadratic
terms $\left( \partial _{i}u_{k}\right) ^{2}$ and $\left( \partial
_{i}\partial _{j}u_{k}\right) ^{2}$, respectively. Whereas the first
(\textquotedblleft piezoelectric") contribution can be understood starting
from the piezoresponse of an isolated planar sheet, the second (
\textquotedblleft flexoelectric") term has the same nature as the
curvature-induced polarization in BN nanotubes, where $P\propto 1/R^{2}$.

It should be stressed that the discovered effect bears no relation to the
formation of the \textit{normal} polarization due to bending of 2D systems
like graphene flat sheets where $P\propto 1/R$ \cite{dumitrica, kalinin}.
From a more general point of view, the effect can be considered as a
mechanically-induced \textit{improper} ferroelectric phase transition, with $%
A$ playing a role of an order parameter. The externally imposed
acoustic-phonon-like displacements $u_{z}(\boldsymbol{r})$ reduces the
initially nonpolar symmetry group to some lower polar subgroup, so that $%
\boldsymbol{P}\left( \boldsymbol{k}\right) $ is expanded in powers of $A$
beginning with the term $\propto A^{2}$. Although $\boldsymbol{P}\left(
\boldsymbol{k}\right) $ is at least quadratic in $A$, the generated
polarization can be very large provided that $A$ is a fraction of the
wavelength $2\pi /k$. The effect may find a variety of practical
applications and, in particularly, for conversion of the ambient wave-like
micromovements into electricity: this tantalizing possibility is discussed
below.

Among the truly 2D crystals that have been recently obtained using the
so-called cleavage technique \cite{novoselov1}, boron-nitride monolayer is
the only piezoelectric with wide band gap. Being partially ionic and
partially covalent with $sp^{2}$-bonding, a flat BN sheet has a remarkable
mechanical flexibility associated with the ease of forming intermediate $%
sp^{2+\alpha }$ bonds. Due to its $D_{3h}$ point symmetry group, it does not
have a ground state polarization, although it exhibits piezoresponse with
the piezoelectric tensor obeying the symmetry relations for the following
non-vanishing components:%
\begin{equation}
e_{x,xx}=-e_{x,yy}=-e_{y,xy}=-e_{y,yx},  \label{eq:piezo}
\end{equation}%
where $x$ is directed along the symmetry axis of 2-th order \cite{landau1}.

To describe periodically and commensurately distorted hexagonal BN sheets,
we introduce the corrugation wave vectors $\boldsymbol{k}$ as $2\pi \mathbf{e%
}/{\lambda }$, where the unit vector $\mathbf{e}$ and wave-length $\lambda $
are expressed via the lattice vectors $\boldsymbol{a}_{1}$ and $\boldsymbol{a%
}_{2}$ of a 2D sheet in the following way: $\mathbf{e}=\boldsymbol{\lambda }%
/\lambda $, with $\boldsymbol{\lambda }(n,m)$ = $n\boldsymbol{a}_{1}+m%
\boldsymbol{a}_{2}$, $\lambda =a\sqrt{n^{2}+nm+m^{2}}$, where $n$ and $m$
are integers, $a_{1}=a_{2}=a$ (Fig. 1). Note that the vector $\boldsymbol{%
\lambda }(n,m)$ coincides with the chiral or circumferential vector defined
in the theory of carbon nanotubes \cite{saito}. The corrugation with the
wave-length $\lambda (n,m)$ leads to a rectangular supercell whose
translational vectors are $\boldsymbol{\lambda }(n,m)$ and some translation
vector $\boldsymbol{T}=N\boldsymbol{a}_{1}+M\boldsymbol{a}_{2}$, which is
perpendicular to $\boldsymbol{\lambda }$. The vector $\boldsymbol{T}$ with
length $T=a\sqrt{N^{2}+NM+M^{2}}$ is completely defined by the vector $%
\mathbf{\boldsymbol{\lambda }}$ \cite{dam}:
\begin{equation}
T=\left\{
\begin{array}{cc}
\sqrt{3}\lambda /p, & N-M\neq 3pq, \\
\sqrt{3}\lambda /\left( 3p\right) , & N-M=3pq,%
\end{array}%
\right.  \label{eq:1}
\end{equation}%
where $p$ is the greatest common divisor of $N$ and $M$, and $q$ is an
integer. It is also convenient to define the vectors $\boldsymbol{k}$ in the
polar coordinate system, ($k,\,\theta $), where $\theta $ is the angle
between the $\boldsymbol{k}$ and $(\boldsymbol{a}_{1}+\boldsymbol{a}_{2})$
\cite{note}. We choose $(\boldsymbol{a}_{1}+\boldsymbol{a}_{2})$ in such a
way that the shortest interatomic distance along this vector is passed from
B to N atom (not from N to B, Fig. 1). Borrowing the terminology from the
theory of nanotubes, we shall call the sine-wave distortions with $\theta
=0\pm \pi l/3$ the \textquotedblleft armchair"-like and with $\theta =\pi
/6\pm \pi l/3$ the \textquotedblleft zigzag"-like, where $l$ is an integer.
All other distortions falling between the two will be called
\textquotedblleft chiral".

\begin{figure}[tbp]
\centering
\includegraphics[height=65mm]{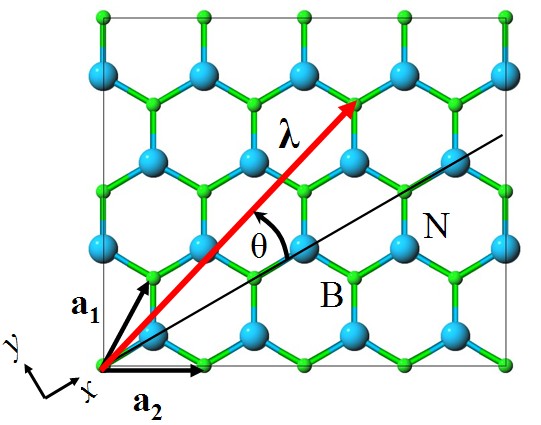}
\caption{Honeycomb BN lattice with basis vectors $\boldsymbol{a}_{1}$ and $%
\boldsymbol{a}_{2}$. The $x$ axis is chosen to be parallel to $(\boldsymbol{a%
}_{1}\boldsymbol{+a}_{2})$, whereas the $y$ axis perpendiculat to it. The
angle between the $(\boldsymbol{a}_{1}\boldsymbol{+a}_{2})$ and the chiral
vector $\boldsymbol{\protect\lambda }$ is the polar angle $\protect\theta $
used in the paper to describe the the corrugation wave vectors $\boldsymbol{k%
}$=$2\protect\pi /\boldsymbol{\protect\lambda }(n,m)$. The example
corresponds to the case of $n=3,m=1$. }
\label{fig:fig1}
\end{figure}

\begin{figure}[tbp]
\centering
\includegraphics[height=65mm]{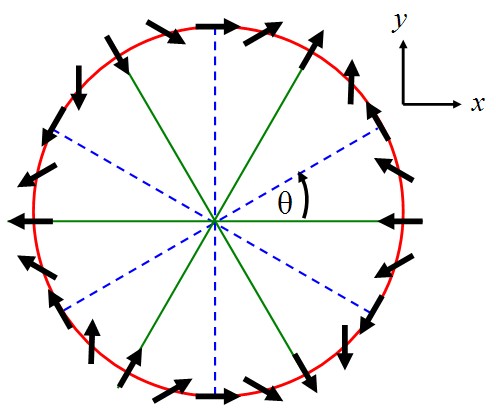}
\caption{A scheme showing the angular dependency of the polarization
(depicted as arrows) induced by sinusoidal corrugations at a fixed
"amplitude-to-wave-length" ratio $Ak$. While the $\boldsymbol{k}$-vector
scans the angle $\protect\theta $ from 0 to 2$\protect\pi $, the Cartesian
components of the polarization change approximately as ($-\cos 2\protect%
\theta $, $\sin 2\protect\theta $), where $\protect\theta $ is the angle
defined in Fig. 1. Green solid and blue dashed lines indicate the armchair
and zigzag directions along which the polarization is pure longitudinal and
transverse, respectively.}
\label{fig:fig2}
\end{figure}

To calculate the corrugation-induced polarization, we use the Berry-phase
approach \cite{vanderbilt1}. Namely, we treat the electronic
polarization as a geometrical phase accumulated by the occupied electrons in
the slow (adiabatic) process of corrugation: $\boldsymbol{P}$ = $\boldsymbol{%
P}^{(u)}$- $\boldsymbol{P}^{(0)}$, with
\begin{equation}
\boldsymbol{P}^{(u)}=-\frac{2ie}{(2\pi )^{2}}\sum_{n=1}^{M}\int_{BZ}d^{2}%
\boldsymbol{k}\left\langle u_{n\mathbf{k}}^{(u)}|\nabla _{\mathbf{k}}u_{n%
\mathbf{k}}^{(u)}\right\rangle ;  \label{eq:2}
\end{equation}%
where $u_{n\mathbf{k}}^{(u)}$ is the periodic part of the Bloch functions
and $e$ is the electron charge. Note that the integral in this expression is
taken over 2D Brillouin zone, so that the value of polarization has a
dimension of a charge per unit length (electron charge per Bohr radius, $%
e/a_{B}$, if atomic units are used). Alternatively, in the simplest case of
isolated bands, the formula (\ref{eq:2}) can be cast as

\begin{equation}
\boldsymbol{P}^{(u)}=-\frac{2e}{S^{(u)}}\sum_{n=1}^{M}\boldsymbol{r}%
_{n}^{(u)},  \label{eq:3}
\end{equation}%
where $\boldsymbol{r}_{n}$ is the center of the Wannier function (WF)
corresponding to the occupied band $n$ and $S^{(u)}=\lambda \,T$ the surface
area. Since the ionic contribution $(e/S^{(u)})\sum_{i}Z_{i}\boldsymbol{R}%
_{i}^{(u)}$ to the total in-plane polarization does not change under the
corrugations of interest, it can be completely neglected.

The integrals $\boldsymbol{P}^{(u)}$ (3) were computed  using  ABINIT  code \cite{abinit}  with  a 12$\times$4$\times$1  Monkhorst-Pack $\mathbf{k}$-point grid, where the largest number 12  corresponds to the direction of  undulations $\boldsymbol{\lambda}$  and 4 to the vector $\boldsymbol{T}$, perpendicular to $\boldsymbol{\lambda}$. The flat and corrugated BN sheets were simulated by a slab-supercell approach with the interplanar  distances of 20$a_{B}$  and larger to ensure negligible wave function  overlap between the replica sheets. For the plane-wave expansion of the valence and conduction band wave-functions, a cutoff energy was used in the range of 80-100 Ry, depending on the corrugation wave length $\lambda $.  The WFs and their geometrical centers $\boldsymbol{r}%
_{n}^{(u)}$ [see (4)] have been calculated  with the PWSCF package \cite{pwscf} in three steps. First, we performed  a self-consistent ground-state calculation and  then  a non-self-consistent calculation  at fixed $\mathbf{k}$-points  for all the occupied  bands. And finally, the obtained Bloch functions were transformed into maximally localized WFs (MLWFs) as described in Ref. \cite{mostofi}. We used $\pi$ and $\sigma$ orbitals as initial guess projection functions allowing to obtain converged result for MLWFs . There are $\mathcal{N}$  $\pi$-derived and 3 $\mathcal{N}$    $\sigma$-derived WFs, where  $\mathcal{N}$  is the number of B-N pairs per unit cell $\lambda \times$T.

Before describing the results of ab-initio modeling, we shall derive a phenomenological
expression for $\boldsymbol{P(k})$ considering the BN sheet as a continuous
elastic membrane, which appears to be very accurate. Obviously, in the long-wave length limit ($\boldsymbol{k}%
\rightarrow 0$) the induced polarization can be expressed via 2D
piezoelectric tensor components $e_{i,jk}$ and pure shear deformations,
because isotropic distortions do not contribute to the polarization of a
flat BN. Substituting the displacement field $u_{z}\left(
\boldsymbol{r}\right) =A\sin (\boldsymbol{k\cdot r}+\varphi )$ into the
standard deformation tensor of a membrane \cite{landau2}
\begin{equation}
\varepsilon _{ij}=\frac{1}{2}\left( \frac{\partial u_{i}}{\partial x_{j}}+%
\frac{\partial u_{j}}{\partial x_{i}}+\frac{\partial u_{z}}{\partial x_{i}}%
\frac{\partial u_{z}}{\partial x_{j}}\right) ,  \label{eq:4}
\end{equation}%
where $x_{i}=(x,y),$ one can easily find the net shear components $\eta
=(\varepsilon _{xx}-\varepsilon _{yy})/2$ and $\gamma =\varepsilon
_{xy}=\varepsilon _{yx}$ as
\begin{eqnarray}
\left(
\begin{array}{c}
\eta  \\
\gamma
\end{array}%
\right) =\frac{\varepsilon_{\parallel}}{2}\left(
\begin{array}{c}
\cos 2\theta  \\
\sin 2\theta
\end{array}%
\right) ,\;
 \label{eq:6}
\end{eqnarray}
where $\varepsilon _{\parallel }=A^{2}k^{2}/4$ is the net film stretch along
the $\boldsymbol{k}$ associated with the third term in Eq.~(\ref{eq:4}) and $%
\theta $ is is the angle related to $\boldsymbol{k}$ the way shown in
Fig.~1. The Cartesian components of polarization in the long wavelength
approximation become $P_{x}=2\alpha \eta ,~P_{y}=-2\alpha \gamma ,$ where $%
\alpha $ is the `clamped-ion' piezoelectric constant $e_{x,xx}$ of a flat BN
sheet. Using these components and relation (\ref{eq:6}), the vector of polarization
can be presented as
\begin{equation}
\boldsymbol{P}\left( \boldsymbol{k}\right) =2\varepsilon _{\parallel }(%
\mathbf{e}_{\parallel }\cos 3\theta -\mathbf{e}_{\perp }\sin 3\theta ),
\label{eq:3theta}
\end{equation}%
where $\mathbf{e}_{\parallel }$ and $\mathbf{e}_{\perp }$ are the unit
vectors parallel and perpendicular to $\boldsymbol{k}$ ($\mathbf{e}_{\perp }$
are defined so that $\boldsymbol{z}\cdot \left[ \mathbf{e}_{\perp }\times
\mathbf{e}_{\parallel }\right] >0$, $\boldsymbol{z}\mathbf{\parallel }\left[
\boldsymbol{x}\times \boldsymbol{y}\right] $). It is clear that the vector $%
\boldsymbol{P(k)}$ simply rotates with $\theta $ for the constant $%
\varepsilon_{\parallel }$.

Being accurate in the long-wave length limit ($\boldsymbol{k}\rightarrow 0$%
), the formula (\ref{eq:3theta}) must be corrected for shorter wavelengths
by supplementing the terms $\propto k^{4}.$ The first such term comes from
the fact that the longitudinal component of polarization is tangent to the
surface and tilted down and up relative to the initial ($x,y$)- plane. It,
therefore, should be projected onto the plane, which has not been taken into
account in Eq. (\ref{eq:3theta}). To obtain this correction, one should
average over $\lambda $ the quantity $\frac{1}{2}\alpha A^{2}k^{2}\mathbf{e}%
_{\parallel }\cos 3\theta \cos ^{2}(\boldsymbol{k\cdot r}+\varphi )\left[
\cos \vartheta (\mathbf{r})-1\right] $, where $\vartheta (\mathbf{r})$ is
the tilting angle. Noticing that $\cos \vartheta (\mathbf{r})-1\approx -%
\frac{1}{2}A^{2}k^{2}\cos ^{2}(\boldsymbol{k\cdot r}+\varphi )$ one can
easily find the average as $-\frac{3}{32}\alpha A^{4}k^{4}\,\mathbf{e}%
_{\parallel }\cos 3\theta $.

The second correction accounts for the flexoelectric effect, i.e. the
appearance of the macroscopic polarization due to finite curvature like in
BN nanotubes. By symmetry, such a polarization is maximal in amplitude for the zigzag and
minimal (zero) for the armchair directions; this angular dependence can be
simply described by $\sin 3\theta $. Besides, the polarization should be
parallel to $\mathbf{e}_{\perp }$, in close analogy with the tubes where the
macroscopic polarization is always along the tube. And, finally, it should
be proportional to the net inverse radius of curvature squared, $\overline{%
1/R^{2}}$, instead of $1/R^{2}$ in the tubes. Defining the principal local
curvature along the $\boldsymbol{k}$ as $\nabla^{2}u_{z}(\mathbf{r})=A\,k^{2}\sin (\boldsymbol{k\cdot r}+\varphi
)=1/R(\boldsymbol{r})$, one can easily find that $%
\overline{1/R^{2}}\equiv \overline{R^{-2}}=A^{2}k^{4}/2$. Thus, the
curvature-induced correction can be represented as $\beta \,\overline{R^{-2}}%
\mathbf{e}_{\perp }\sin 3\theta $, where $\beta $ is the flexoelectric
constant.

By adding the above corrections to the expression (\ref{eq:3theta}), we come
to the final result:
\begin{eqnarray}
\boldsymbol{P}\left( \boldsymbol{k}\right) &=&\alpha \varepsilon _{\parallel
}\left[ \mathbf{e}_{\parallel }\left( 1-\frac{3}{2}\varepsilon _{\parallel
}\right) \cos 3\theta -\mathbf{e}_{\perp }\sin 3\theta \right]  \notag \\
&&+\beta \overline{R^{-2}}\mathbf{e}_{\perp }\sin 3\theta .  \label{eq:8}
\end{eqnarray}%
 This formula is invariant with respect to rotation $%
\theta \rightarrow \theta +2\pi /3,$ as one would have expected from the $%
D_{3h}$ symmetry. The first term comes from the increase in surface area that accompanies the
sine-wave atomic displacements, it represents the
piezoelectric effect. The second term, on the contrary, is non-zero only due to
finite averaged curvature squared, obviously, it renders the flexoelectric
effect. Since $\alpha <0$ and $\beta >0$, the extrema of the polarization
are:%
\begin{eqnarray}
P_{\max } &=&|\alpha |\varepsilon _{\parallel }+\beta \overline{R^{-2}}%
,\quad \,\,\,\quad \text{zigzag,}  \label{eq:10} \\
P_{\min } &=&|\alpha |\varepsilon _{\parallel }\left( 1-\frac{3}{2}%
\varepsilon _{\parallel }\right) ,\quad \text{armchair.}
\end{eqnarray}%
For all the other (chiral) directions, the amplitude falls within the above
limits.

The calculations show that for relatively small $A/\lambda \lesssim 0.2$ the
dependence $\boldsymbol{P(k})$ does follow the continuous expression (\ref%
{eq:8}) very well. The mutual orientation of the $\boldsymbol{P}$ and $%
\boldsymbol{k}$ constantly evolves as the latter rotates through the angle $%
\theta $ (Fig. \ref{fig:fig2}). $\boldsymbol{P}$ becomes pure longitudinal
along the armchair $\boldsymbol{k}-$directions and pure transversal along
the zigzag directions. Along the chiral $\boldsymbol{k}$-directions, it has
both components. Being practically independent of the initial phase $\varphi
$, the magnitude of polarization $P$ slightly depends on $\boldsymbol{k}$
(within $30\%$) and reaches its minima along the armchair and maxima along
the zigzag directions.

We have found $\alpha $ and $\beta $ numerically to be $-0.118e/a_{B}$ and $%
0.044ea_{B}$, respectively. To check that the flexoelectric constant $\beta $
is indeed related to the properties of BN nanotubes, we considered a set of
`ideal' zigzag ($n,0$) nanotubes cut out of BN sheets in such a way that the
B and N ions fall on the cylinder with radius $R$, $2\pi R$ coinciding with
the `wave-length' $\lambda (n,0)$. These tubes are \emph{longitudinally}
polarized and their polarization is proportional to $1/R^{2}$ to a leading
order in $1/R$. The coefficient of proportionality, as we found from
additional calculations, is $0.042ea_{B}$, which is very close to the
parameter $\beta $ ($0.044ea_{B}$). This means that the curvature-induced
polarizations in the corrugated BN sheets and in the BN nanotubes indeed
have the same origin.

\begin{figure*}[tbp]
\centering
\includegraphics[scale=1.25]{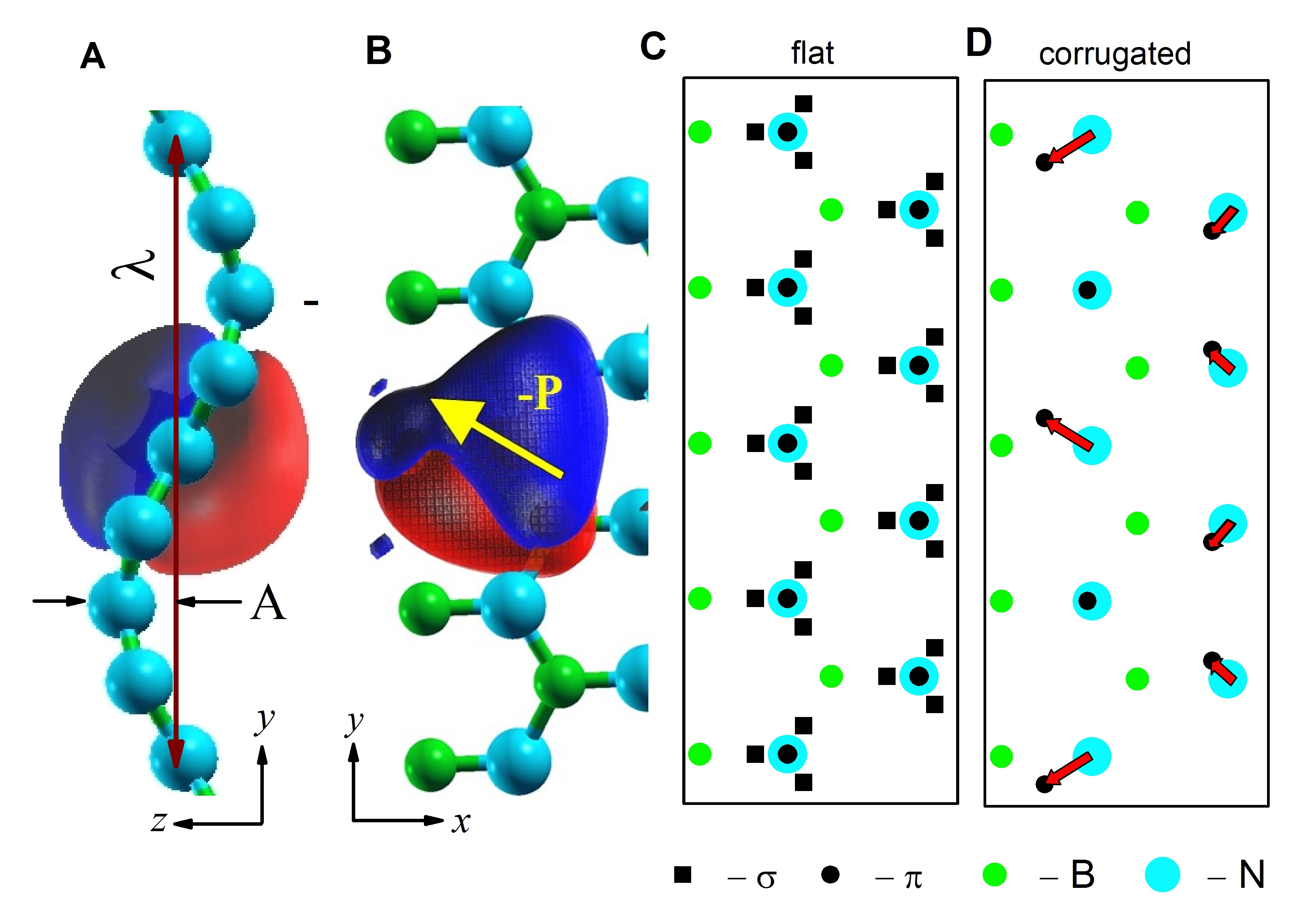}
\caption{Wannier functions and their shifts in a corrugated BN sheet
corresponding to a zigzag period vector $\boldsymbol{\protect\lambda }(4,-4)$ $%
(\protect\theta =90^{\circ })$, $\ A/\protect\lambda $=0.1, phase $\protect%
\varphi $= 0. ($\mathbf{A}$ and $\mathbf{B)}$ Side and top views of a $%
\protect\pi $-like WF contributing most to the polarization (isosurface $%
=\pm 0.9$). The yellow arrow indicate the shift of the WF in the $(x,y)$
plane, which is antiparallel to the locally induced polarization $\mathbf{P}$%
. ($\mathbf{C}$ and $\mathbf{D}$) The $(x,y)$- positions of the centers of
the $\protect\pi$ and $\protect\sigma$ WFs in a flat and corrugated BN
sheet, respectively. The positions of the centers of $\protect\sigma$ WFs
are indicated by black squares, while those of $\protect\pi $ by black
circles. In the initial flat state, the centers of $\protect\pi$ WFs
coincide with those of N atoms. The red arrows show the most pronounced
shifts in $\protect\pi$ WFs induced by the corrugation (the amplitudes of
the shifts are tripled for clarity). Note that the yellow arrow points along
one of the red ones.}
\end{figure*}
\begin{figure*}[tbp]
\centering
\includegraphics[scale=0.6]{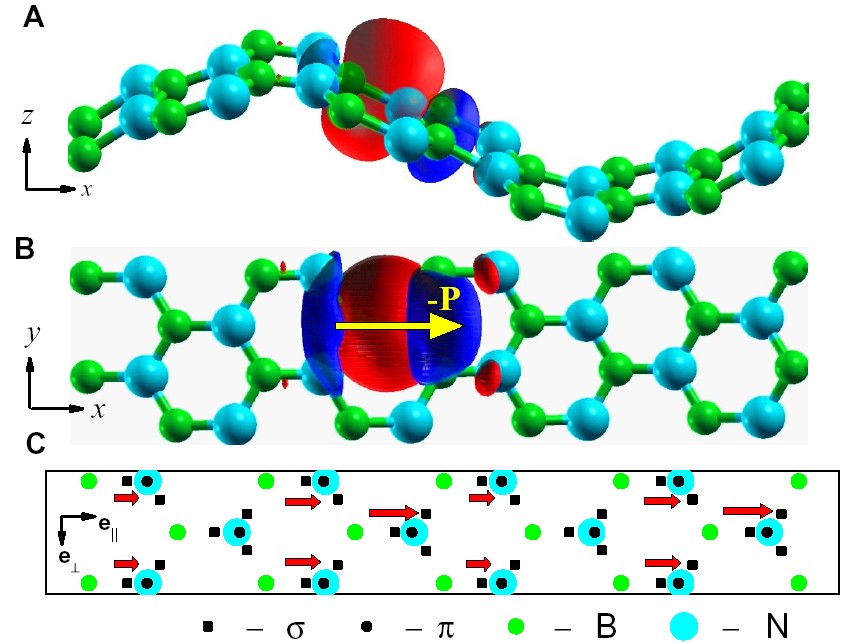}
\caption{Wannier functions and their shifts in a corrugated BN sheet
corresponding to a armchair period vector $\boldsymbol{\protect\lambda }(4,4)$ $(%
\protect\theta =0^{\circ })$, $A/\protect\lambda $=0.1, $\protect\varphi $=
0. ($\mathbf{A}$ and $\mathbf{B)}$ Side and top views of a $\protect\sigma $%
-like WF contributing most to the polarization (isosurface $=\pm 0.9$). The
yellow arrow indicates the shift of the WF in the $(x,y)$ plane, which is
antiparallel to the locally induced polarization $\mathbf{P}$. The $(x,y)$-
positions of the centers of the $\protect\pi $ and $\protect\sigma $ WFs in
the corrugated BN sheet ($\mathbf{C}$). The positions of the centers of $%
\protect\sigma $ WFs are indicated by black squares, while those of $\protect%
\pi $ by black circles. The red arrows show the most pronounced shifts in $%
\protect\sigma$ WFs induced by the corrugation (the amplitudes of the
shifts are scaled up for clarity). Note that the yellow arrow corresponds to
the red one.}
\end{figure*}

To analyze the relationship between electronic structure and $\boldsymbol{P}%
\left( \boldsymbol{k}\right) $, we notice that the WF in a flat BN sheet
have roughly the character of the ${\sigma }$- and ${\pi }$-bond orbitals (${%
\sigma }$ and ${\pi }$ WFs). The $\pi -$like WFs are centered exactly on the
N atoms, whereas their $\sigma $ counterparts are somewhere in the middle of
the B-N bonds closer to the N atoms (Fig. 3C). The contributions of the $\pi
$ and $\sigma $ WFs to the `clamped-ion' polarization induced by an in-plane
uniaxial stretching have the \textit{same} sign, with the $\pi $ WFs
dominating and amounting to as much as 85$\%$ of the total (see also Ref.
\cite{sai}).

Although the corrugation of a flat BN sheet leads to the rehybridization
effects like mixing of $\pi $ and $\pi ^{\ast }$ electronic bands due to
breaking the $z\rightarrow -z$ mirror symmetry \cite{kim}, it is still
possible to classify all the WFs into $\pi $ and ${\sigma }$ types, Figs.~3
and 4. It might naively appear that in the corrugated structures the main
contribution to $\boldsymbol{P}\left( \boldsymbol{k}\right) $ would also
come from the $\pi $ WFs. In fact, this is not the case. The contribution of
$\sigma $ WFs generally increases, becoming strongly dominant in some
special cases in corrugated films.

Consider first the case of zigzag corrugation with $n=4$, $m=-4$, $A/\lambda
=0.1$ and $\varphi =0$, Fig.~3. Here, the shifts of $\pi$ WFs are still
dominant but not so apparently and $|P_{\pi }/P_{\sigma }|$ is only about $%
2.5$. In contrast to the flat case, the $\pi$ and $\sigma$ WFs have the
\textit{opposite} sign contributions to $\boldsymbol{P}(\boldsymbol{k})$
exhibiting \textit{inhomogeneous} shifts within the two-dimensional
supercell ${\lambda }\times T$. Upon the corrugation, the $\pi$ WF centers
shift relative to the initial positions as shown in Fig. 3D. Although the
sum of the shifts along the $y$ axis cancels within the supercell, the
similar sum along the $x$ axis is nonzero, which means the appearance of a
macroscopic transverse polarization, $\boldsymbol{P}\perp \boldsymbol{k}$.
Interestingly, the magnitude of the shifts is larger in flatter and more
stretched areas, and the shifts themselves (along $y$) are always directed
towards more curved and less stretched areas. Because of the charge transfer
from the flatter to curved regions, the conduction band maximum increases
and the valence minimum increases in the curved areas, thus leading to a
space modulation of the band gap with the wavelength $\lambda $.

Now, we turn our attention to another case, represented by the armchair
undulation with $n=4$, $m=4$, $A/\lambda =0.1,$ and $\varphi =0$, Fig. 4.
Here again, the $\pi $ and $\sigma $ WFs give the \emph{compensating}
contributions to the polarization. But now, in contrast to the zigzag case, $%
\sigma$ WFs dominate by a wide margin: $|P_{\sigma }/P_{\pi }|\sim 4.5$.
All the WFs shift strictly either along or opposite the $\boldsymbol{k}$%
-vector (Fig. 4C), so that the induced polarization is purely longitudinal.
The dominant ${\sigma }$ WFs move towards the N-atoms, thus further
increasing the polarity of the ${\sigma }$ B-N bonds. The magnitude of the
shifts, as in the previous case, is larger in more flattened and stretched
areas, but this works only in the second and fourth quarters of the $\lambda
$. In the first and third quarters, the shifts of ${\sigma }$ WFs are modest
even in the well stretched areas. The charge transfer, therefore, increases
the energy gap mainly in the second and fourth quarters at the expense of
the first and third quarters of the  wavelength.

Taking a reasonable amplitude-to-wavelength ratio of $A/\lambda \sim 0.2$,
it is easy to estimate from Eq.~(\ref{eq:8}) that the induced polarization
can be on the order of $4\times 10^{-2}e/a_{B}$. This quantity cannot be
compared directly to that of 3D bulk materials because the latter has
different dimension, $e/a_{B}^{2}$. The comparison can be performed,
nevertheless, if both 2D and 3D polarizations are expressed as the total
dipole per stoichiometric unit \cite{sai}; then the polarization estimated
above becomes $0.8ea_{B}$. It is tempting to compare this value with that in
the standard perovskite PbZr$_{x}$Ti$_{1-x}$O$_{3}$ that has very high
efficiency for converting electricity into mechanical strain and vice versa.
Taking for the latter an experimental \textquotedblleft bulk" piezoelectric constant $e_{33}$
of $11.3$ C/m$^{2}$ \cite{wu} and an achievable strain of $0.2\%$ \cite{park}%
, we obtain only $0.2ea_{B}$, \emph{four} times smaller than the one for BN
(!). Hence, the generated polarization in a BN sheet can be comparable or
even higher than that in best perovskite ferroelectrics. This situation
pretty much resembles that in the field of electrostrictive materials.
Usually, the electrostriction is a small effect, because the induced
deformation is only \textit{quadratic} in electric field or polarization.
However, dielectrics with large polarization, such as the relaxor
ferroelectric Pb(Mg$_{1/3}$Nb$_{2/3}$)O$_{3}$-PbTiO$_{3}$, are capable of
exhibiting the exceptionally large electrostrictive strains.

The important characteristic, the voltage drop $U\ $between opposite sides
of the BN strip, can be estimated as ($P/\pi \epsilon \epsilon _{0})\ln (L/b)
$, where $P$ is the usual 2D polarization (in units of C/m), $L$ the
separation between the charged strip edge states (strip width), and $b$ the
effective radius of those states, $\epsilon _{0}$ the dielectric
permittivity of vacuum. Taking $b$ to be on the order of the lattice
parameter $a$, for reasonable $L=(100-1000)a,$ and $P=10^{-2}$ $e/a_{B},$ we
formally obtain $U\lesssim 5-8$V for BN strip suspended in vacuo and smaller
values for the strip on dielectric substrate like SiO$_2$ ($\epsilon =3.9)$, or
more than 20 times that of ZnO nanowires \cite{schubert}. Interestingly, in
case of suspended BN the band bending $qU$ exceeds the value of the bandgap $%
E_{g}=5.8$eV \cite{ng}, meaning that it will result in a charge transfer between the charged edges
to make $qU\approx E_{g}$. This would mean that the achieved bias voltage would
not depend on the strip width. For smaller bare values of the bias, it
depends on $L$ only logarithmically, i.e. very weakly. This insensitivity of
the produced bias voltage is extremely attractive for applications. The
ability of a BN sheet to effectively convert wave-like deformations into
electricity makes this material very promising for electromechanical
applications at the nanoscale (Fig.~5). Below, we present a new concept of a
nanogenerator powered by an ambient motion or agitation.
\begin{figure}[tbp]
\centering
\includegraphics[height=35mm]{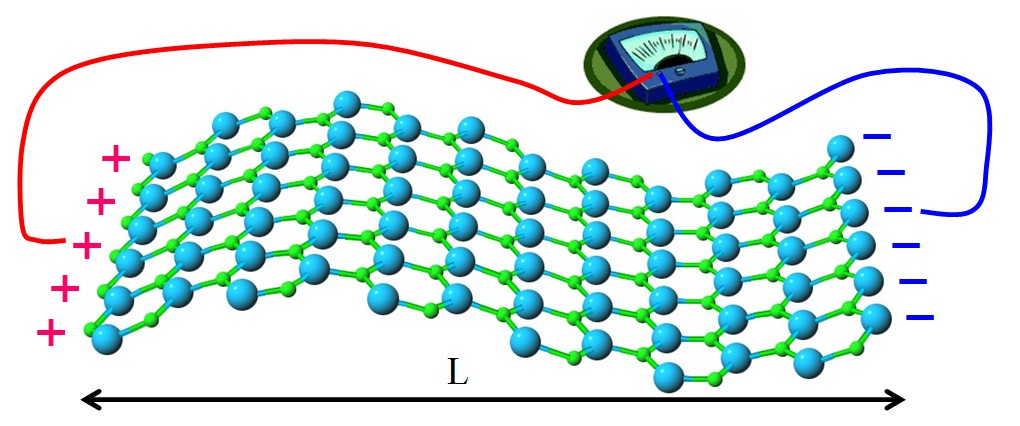}
\caption{Schematic showing generation of a bias voltage between the edges of
a corrugated BN nanosheet (the geometry corresponds to Fig. 4). Note
that the induced polarization is longitudinal, because the corrugation is of
the armchair type (see text).}
\label{fig:fig5}
\end{figure}

The direct and efficient conversion of mechanical energy into electricity is
of great interest for practical applications. Recently, much attention has
been attracted to the possibility of creating miniature nano-generators
using semiconducting piezoelectric nanowires such as ZnO, GaN, and CdS \cite%
{wang1, wang2, qin}; during bending they generate the voltage bias between
the opposite sidewalls. Unfortunately, such nanowires are not really
suitable for this application. First, as has been pointed out in Ref.~\cite%
{schubert}, they lose a significant fraction of their generated charges
already in the process of bending due to large internal conductivity by
virtue of being semiconducting. Second, since the nanowires are grown
randomly, only a small fraction of them (only about $1\%$) contributes to
the generated electric current. And third, in order to generate a \textit{dc}
current, the nanowire should slide over a metallic electrodes until the
nanowire/electrode interface becomes a forward-biased Schottky barrier; this
is equivalent to a \textit{ac-dc} converter.

The nanogenerator based on BN sheets does not suffer from all of the above
shortcomings. First, the BN sheet is a dielectric nanostructure with a
relatively wide energy gap ($\sim 5.8$ eV). Second, all the atoms in the
sheet are involved into producing electricity (it is the \textquotedblleft bulk" effect).
Third, the nanogenerator does not need any \textit{ac-dc} converter, like a
diode bridge rectifier, because the generated polarization has a \emph{%
constant} direction provided that the direction of the wave vectors $%
\boldsymbol{k}$ is fixed or, in other words, the ambient wave-like movements
are oriented in the space. Fourth, the produced bias voltage is nearly
independent from the BN strip width. Moreover, all these advantages couple
with outstanding mechanical properties of a BN sheets, like their ability to
be stretched or bent by several percent with no atomic defects involved \cite%
{saito}.

It should be stressed that the corrugation-induced polarization can not be
observed in a thin BN film cut out of the hexagonal bulk BN material. In
such a film, the B and N arrangement is reversed in the neighboring
hexagonal atomic layers, so that the polarization produced by each such pair
of layers cancels out. However, the effect can be observed in an ultra-thin
BN films with \emph{odd} number of layers, 1, 3, 5, ... Here, of all the
layers only one is going to produce the polarization, while the rest can be
considered as a mechanically supporting substrate or scaffold.

It is worth mentioning that the BN monolayer is not the only 2D system
capable of generating electricity under periodic bending. Its isoelectronic
analog BC$_{2}$N exhibits similar effect in the ground state, like, for
example, structure found with the LDA total energy calculations \cite{liu}.
Due to absence of inversion symmetry, BC$_{2}$N in this structure is already
polarized (electret)\ in the initial flat state with a polarization of $%
0.73\times 10^{-2}\,e/a_{B}$ pointing along the armchair-type direction
perpendicular to the pure C-C chains. Moreover, due to in-plane anisotropy
under $2\pi /3$ rotations, the corrugation-induced polarization $\boldsymbol{%
P}\left( \boldsymbol{k}\right) $ is no longer described by the formula (\ref%
{eq:8}) and its magnitude depends on $\theta $ much stronger than that in
BN. Nevertheless, there are some $\boldsymbol{k}$-directions along which the
generation of polarizations is practically as effective as in BN, such
directions include the parallel and perpendicular ones to the
above-mentioned C-C chains.

Here, we assumed that the edges of BN sheets are fixed and, therefore, the
surface area of the membrane increases due to its deformation in the \textit{%
third} dimension, and this increase is simply $\varepsilon _{\parallel }$.
It is possible, however, to imagine the situation when the edges are not
clamped and the layer bends in accordion-like fashion without any in-plane
(average) stretching. In this case, the in-plane deformations $\partial
_{x}u_{y}$ and $\partial _{y}u_{x}$ would cancel the nonlinear term $%
\partial _{x}u_{z}\partial _{y}u_{z}$ in Eq.~(\ref{eq:4}), and the induced
polarization would be defined \emph{entirely}\textit{\ }by the second,
`flexoelectric' term in (\ref{eq:8}). Such a polarization, however, is
relatively modest because the elastic energy associated with the
out-of-plane bending is noticeably smaller than the elastic stretching
energy \cite{sanchez} (this, of course, is already evident from the
comparison of the parameters $\alpha $ and $\beta $).

We have studied above the effectively infinite BN sheets satisfying the
periodic boundary conditions, and the important question is whether the
present results apply to the finite 2D BN systems. This question can be
motivated by an example of a finite graphene sheet where the presence of
zigzag edges induces the localized electronic states at each edge, thus
modifying the properties of the initial infinite system \cite{nakada}. Shall
we expect the same effect in the case of BN? The answer is no, because,
contrary to the carbon counterpart, the BN sheet does not have the
conduction and valence bands that touch in a linear fashion at two
inequivalent Dirac points, $\boldsymbol{K}$ and $\boldsymbol{K}^{\prime }$%
. This degeneracy is always lifted in a 2D BN due to a broken sublattice
symmetry, so that the system remains insulating. In the present case the
difference between $\boldsymbol{P}$ in flat and electroded corrugated states
in finite systems can be calculated by using periodic boundary conditions
\cite{vanderbilt}.

In summary, we have found that there is a strong polarization-corrugation
coupling in a two-dimensional strip cut out of BN sheet, which can be
decomposed into the nonlinear `piezoelecric' and the flexoelectric effects.
The direction of the induced polarization strongly depends on the
corrugation wave vector $\boldsymbol{k}$ and changes approximately as ($%
-\cos 2\theta $, $\sin 2\theta )$, where $\theta $ is the angle between the
zigzag axes and $\boldsymbol{k}$. The magnitude of the polarization as a
total dipole moment per chemical formula can reach very large values
comparable to those in the best known perovskite piezoelectrics. The present
effect may be used in nanoscale generators activated by  ambient
vibrations that can power small electronic devices and circuits.


\begin{thebibliography}{99}
\bibitem{martin} {\small \ R. Martin, Phys. Rev. B \textbf{5}, 1607 (1972). }

\bibitem{tagantsev} {\small \ A. K Tagantsev, Phys. Rev. B \textbf{34}, 5883
(1986). }

\bibitem{cross} {\small \ L. E. Cross, J. Mater. Sci. \textbf{41}, 53
(2006). }

\bibitem{sharma} {\small \ N. D. Sharma, R. Maranganti, P. Sharma, J. Mech.
Phys. Solids \textbf{55}, 2328 (2007). }

\bibitem{saito} {\small \ R. Saito, G. Dresselhaus, M.S. Dresselhaus, \emph{%
Physical Properties of Carbon Nanotubes} (Imperial College, London, 1998). }

\bibitem{meyer} {\small J. C. Meyer \emph{et al.}, Nature \textbf{446}, 60
(2007). }

\bibitem{mele} {\small \ E.J. Mele, P. Kr\'al, Phys. Rev. Lett. \textbf{88},
056803 (2002). }

\bibitem{nakhamson} {\small S. M. Nakhmanson, A. Calzolari, V. Meunier, J.
Bernholc, M. B. Nardelli, Phys. Rev. B \textbf{67}, 235406 (2003). }

\bibitem{sai} {\small N. Sai, E. J. Mele, Phys. Rev. B \textbf{68}, 241405
(2003). }

\bibitem{dumitrica} {\small T. Dumitric\u{a}, C. M. Landis, B. I. Yakobson,
Chem. Phys. Lett. \textbf{360}, 182 (2002). }

\bibitem{kalinin} {\small \ S. V. Kalinin, V. Meunier, Phys. Rev. B \textbf{%
77}, 033403 (2008). }

\bibitem{novoselov1} {\small K. S. Novoselov \emph{et al.}, Proc. Natl.
Acad. Sci. USA \textbf{102}, 10451 (2005). }

\bibitem{landau1} {\small L. D. Landau, E. M. Lifshitz, \textit{%
Electrodynamics of Continious Media} (Pergamon, New York, 1993). }

\bibitem{vanderbilt1} {\small R. Resta, D. Vanderbilt, Theory of
polarization: A modern approach, in \textit{Physics of ferroelectrics: a
Modern perspective}, ed. by K. M. Rabe, C. H. Ahn, and J.-M. Triscone
(Springer, Berlin, 2007), pp. 31-68. }

\bibitem{abinit} {\small ABINIT code is a common project of the Universite Catholique de
Louvain, Corning Incorporated, and other contributors (URL  http://www.abinit.org).}

\bibitem{pwscf} {\small S. Baroni \emph{et al.},  http://www.pwscf.org.}  

\bibitem{mostofi} {\small A.A. Mostofi et  al.  Comput. Phys.  Commun. \textbf{178}, 685 (2008).}

\bibitem{dam} {\small M. Damjanovi\'{c}, T. Vikovi\'{c}, T. Milo\u{s}evic,
B. Nikoli\'{c}, Acta Cryst. \textbf{A57}, 304 (2001). }

\bibitem{note} {\small Note that the introduced angle $\theta $ is rotated
by $30^{\circ }$ with respect to the usual chiral angle defined in the
theory of carbon nanotubes, Ref. \cite{saito}. }

\bibitem{landau2} {\small L. D. Landau, E. M. Lifshitz, \textit{Theory of
Elasticity} (Butterworth-Heinemann, Oxford, 1995). }

\bibitem{kim} {\small Y.-H Kim, K. J. Chang, S. G. Louie, Phys. Rev. B
\textbf{63}, 205408 (2001). }

\bibitem{sanchez} {\small D. S\'anchez-Portal, E. Hern\'andez, Phys. Rev. B
\textbf{66}, 235415 (2002). }

\bibitem{wu} {\small Z. Wu, H. Krakauer, Phys. Rev. B \textbf{68}, 014112
(2003). }

\bibitem{park} {\small S.-E. Park, T. R. Shrout, J. Appl. Phys. Cryst.
\textbf{82}, 1804 (1997). }

\bibitem{zhang} {\small Y. Zhang, Y-W. Tan, H. L. Stormer, P. Kim, Nature
\textbf{438}, 201 (2008). }

\bibitem{wang1} {\small Z. L. Wang, J. Song, Science \textbf{312}, 242
(2006). }

\bibitem{wang2} {\small X. Wang, J. Song, J. Li, Z. L. Wang, Science \textbf{%
316}, 102 (2007). }

\bibitem{qin} {\small Y. Qin, X. Wang, Z. L. Wang, Nature \textbf{451}, 809
(2008). }

\bibitem{schubert} {\small M. A. Schubert, S. Senz, M. Alexe, D. Hesse, U. G%
\"{o}sele, Appl. Phys. Lett. \textbf{316}, 122904 (2008). }

\bibitem{ng} {\small M.-F. Ng, R. Q. Zhang, Phys. Rev. B \textbf{69}, 115417
(2004). }

\bibitem{liu} {\small A. Y. Liu, R. Wentzcowitch, M.L. Cohen, Phys. Rev. B
\textbf{39}, 1760 (1989). }

\bibitem{nakada} {\small K. Nakada, M. Fujita, G. Dresselhaus, M. S.
Dresselhaus, Phys. Rev. B \textbf{54}, 17954 (1996). }

\bibitem{vanderbilt} {\small D. Vanderbilt, R. D. King-Smith, Phys. Rev. B \textbf{%
48}, 4442 (1993). }
\end{thebibliography}
\end{document}